\documentclass[journal=jacsat,manuscript=article]{achemso}
\setkeys{acs}{articletitle = true}

\usepackage[version=3]{mhchem} % Formula subscripts using \ce{}
\usepackage[T1]{fontenc}       % Use modern font encodings
\usepackage{xkeyval}
\usepackage{caption}
\usepackage{float}
\usepackage{geometry}
\usepackage{natbib}
\usepackage{setspace}
\usepackage{epstopdf}
\usepackage{amssymb}
\usepackage{ulem}  % for strikethrough the text  (Anna)
\usepackage{siunitx} %per unit√† di misura sistema internazionale
\usepackage{subfig}
\usepackage{physics}
\usepackage{mhchem}

\author{Sofia Pazzagli}
\affiliation{Dipartimento di Fisica ed Astronomia, Universit\`a di Firenze, Via Sansone 1, I-50019 Sesto F.no, Firenze, Italy}
\alsoaffiliation{CNR-INO, Istituto Nazionale di Ottica, Via Carrara 1, 50019 Sesto F.no, Firenze, Italy}
\email{sofia.pazzagli@unifi.it}
\author{Pietro Lombardi}
\affiliation{CNR-INO, Istituto Nazionale di Ottica, Via Carrara 1, 50019 Sesto F.no, Firenze, Italy}
\alsoaffiliation{LENS and Universit\`a di Firenze, Via Carrara 1, 50019 Sesto F.no, Firenze, Italy}
\author{Daniele Martella}
\affiliation{LENS and Universit\`a di Firenze, Via Carrara 1, 50019 Sesto F.no, Firenze, Italy}
\author{Maja Colautti}
\affiliation{LENS and Universit\`a di Firenze, Via Carrara 1, 50019 Sesto F.no, Firenze, Italy}
\author{Bruno Tiribilli}
\affiliation{CNR-ISC Istituto dei Sistemi Complessi, via Madonna del Piano 10, I-50019 Sesto F.no, Firenze, Italy}
\author{Francesco Saverio Cataliotti}
\affiliation{Dipartimento di Fisica ed Astronomia, Universit\`a di Firenze, Via Sansone 1, I-50019 Sesto F.no, Firenze, Italy}
\alsoaffiliation{CNR-INO, Istituto Nazionale di Ottica, Via Carrara 1, 50019 Sesto F.no, Firenze, Italy}
\alsoaffiliation{LENS and Universit\`a di Firenze, Via Carrara 1, 50019 Sesto F.no, Firenze, Italy}
\alsoaffiliation{QSTAR, Largo Fermi 2, I-50125 Firenze, Italy}
\author{Costanza Toninelli}
\affiliation{CNR-INO, Istituto Nazionale di Ottica, Via Carrara 1, 50019 Sesto F.no, Firenze, Italy}
\alsoaffiliation{LENS and Universit\`a di Firenze, Via Carrara 1, 50019 Sesto F.no, Firenze, Italy}
\alsoaffiliation{QSTAR, Largo Fermi 2, I-50125 Firenze, Italy}
\email{toninelli@lens.unifi.it}

\title[nanocrystals]
{Photostable single-photon emission from self-assembled nanocrystals of polycyclic aromatic hydrocarbons}

\abbreviations{IR,UV}

\begin{document}

%%%%%%%%%%%%%%%%%%%%%%%%%%%%%%%%%%%%%%%%%%%%%%%%%%%%%%%%%%%%%%%%%%%%%

% Some journals require a graphical entry for the Table of Contents.
% This should be laid out ``print ready'' so that the sizing of the
% text is correct.

% Inside the \texttt{tocentry} environment, the font used is Helvetica
% 8\,pt, as required by \emph{Journal of the American Chemical
% Society}.

% The surrounding frame is 9\,cm by 3.5\,cm, which is the maximum
% permitted for  \emph{Journal of the American Chemical Society}
% graphical table of content entries. The box will not resize if the
% content is too big: instead it will overflow the edge of the box.

% This box and the associated title will always be printed on a
% separate page at the end of the document.

\begin{tocentry}
\centering
\includegraphics[width=0.85\textwidth]{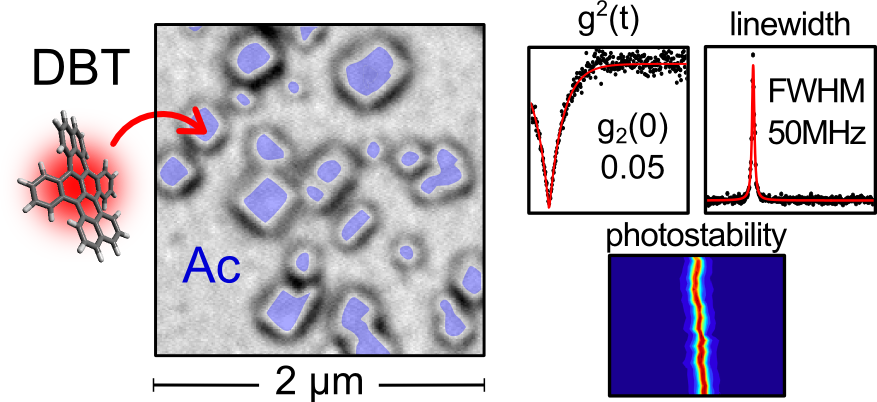}
%\caption{..}
%\label{TOC}
\end{tocentry}

%%%%%%%%%%%%%%%%%%%%%%%%%%%%%%%%%%%%%%%%%%%%%%%%%%%%%%%%%%%%%%%%%%%%%
\begin{abstract}
%ABSTRACT (250 words) + toc (9\,cm by 3.5\,cm)
Quantum technologies could largely benefit from the control of quantum emitters in sub-micrometric size crystals. These are naturally prone to the integration in hybrid devices, including heterostructures and complex photonic devices. Currently available quantum emitters sculpted in nanocrystals suffer from spectral instability, preventing their use as single photon sources \textit{e.g.}, for most quantum optics operations. In this work we report on unprecedented performances of single-photon emission from organic nanocrystals (average size of hundreds \SI{}{\nano\meter}), made of anthracene (Ac) and doped with dibenzoterrylene (DBT) molecules. The source has hours-long photostability with respect to frequency and intensity, both at room and at cryogenic temperature. When cooled down to \SI{3}{K}, the 00-zero phonon line shows linewidth values (\SI{50}{MHz}) close to the lifetime-limit. Such optical properties in a nanocrystalline environment make the proposed organic nanocrystals a unique single-photon source for integrated photonic quantum technologies.

% The intensity autocorrelation function at zero delay source purity measured from the  $g^2(0)$ is as low as 0.05, while detected count rate at saturation amounts to \SI{1.5}{\mega\hertz}.
\end{abstract}

%%%%%%%%%%%%%%%%%%%%%%%%%%%%%%%%%%%%%%%%%%%%%%%%%%%%%%%%%%%%%%%%%%%%%
%\section{Introduction}
%1000 words for an Article (we are at about 500 words)
The ubiquitous deployment of nanocrystals (NCs) in photonics stems from the impressive tunability of their physical and chemical properties, combined with the nano-positioning opportunities offered by support-free colloids and from the possibility of mass-production at low costs \cite{Kovalenko2015}. These features have also promoted NCs as efficient biological markers for imaging \cite{Zheng2015, Lyu2016}, color filters in liquid crystal displays \cite{Kim2013a}, as well as functionalizing elements in light emitting and light harvesting devices \cite{Talapin2010}. 
On the other hand, advanced nanophotonic applications are emerging based on the generation, manipulation and detection of single photons \cite{OBrien2009, Aharonovich2016}. Indeed, leveraging single-photon statistics and quantum coherence for sub-diffraction imaging \cite{GattoMonticone2014}, quantum cryptography\cite{Sangouard2012}, simulation \cite{Tillmann2015}, enhanced precision measurements and information processing \cite{Knill2001} have become roadmap targets for the next 10-20 years \cite{Qmanifesto}. Single-photon sources based on quantum emitters hold promise for these applications because of their on-demand operation \cite{Lounis2005, Chu2017, Loredo2017, Sipahigil2014, Lettow2010a}. However, despite great efforts in the last years to attain controllable sources by coupling solid-state emitters to nanophotonic structures, each platform privileges either the freedom in the device design \cite{Bermudez-Urena2015, Schroeder2011, Liebermeister2014, Riedrich2014, Schell2013, Shi2016} or the quality of single-photon emission \cite{Somaschi2016, Sapienza2015}. Deterministic positioning and control of quantum emitters remains elusive for epitaxial quantum dots \cite{Arcari2014, Daveau2017, Zadeh2016, Davanco2017}, color centers in bulk diamond \cite{Hausmann2012, Mouradian2015} and organic molecules in crystalline matrices \cite{Turschmann2017, Lombardi2017, Checcucci2016, Skoff2016}. On the other hand, versatile approaches based on today-available NCs present important shortcomings with respect to single-photon applications. Photoinduced charge rearrangements in the passivation layer and in the environment of inorganic semiconductors quantum-dot NCs \cite{Pisanello2013, Liu2017} lead to spectral instability of the exciton line \cite{Empedocles1997}, hindering basic quantum optics operations with the emitted photons. Moreover, intermittence in the photoluminescence \cite{Efros2016}, named blinking, seriously affects the average fluorescence quantum yield and hence the photon state purity. Although important results have been obtained by improving synthesis protocols \cite{Chandrasekaran2017} or introducing perovskite materials \cite{Park2015, Raino2016}, the emitter photostability in time or frequency is still below expectations. Notably, similar issues characterize the emission of color centres in nanodiamonds, including those which possess superb optical properties in bulk such as the widely studied negatively charged silicon vacancy \cite{Jantzen2016, Sipahigil2014}, or chromium-related defects \cite{Tran2017}. Hence, despite the wealth of materials and protocols, there still are fundamental limitations for the use of NCs in single-photon applications.
% Quantum technologies could hence largely benefit from the development of tailored NCs emitting coherent photons, one at a time, being easy to process and nanoposition \cite{Livneh2016}. 
 
We here propose and report on self-assembled and support-free organic NCs %/nanoplatelets 
(hundreds \SI{}{\nano\meter} in size) of anthracene doped with single fluorescent dibenzoterrylene molecules (DBT:Ac). We demonstrate that the remarkable features of the bulk system \cite{Toninelli2010a, Nicolet2007, Trebbia2009}, belonging to the family of Polycyclic Aromatic Hydrocarbons (PAH), are preserved in a nanocrystalline environment.
In particular, DBT:Ac NCs exhibit bright and photostable single-photon emission at room temperature that is spectrally stable and almost lifetime-limited (\SI{50}{MHz}) at cryogenic temperatures. The combination of such properties is unique and opens the way to the use of organic NCs for quantum technologies and for single-photon applications in general.

\section{Results and discussion}
We adapted a simple, cost-effective and well-established reprecipitation method \cite{Horn2001, Kasai1992, Kang2004, Baba2011} to grow Ac NCs doped with controlled concentration of DBT molecules (for details see the Experimental Section). In this procedure, a dilute solution of the compounds prepared with a water-soluble solvent (acetone, in our case) is injected into sonicating water where it divides into many droplets. The solvent gradually dissolves and correspondingly the concentration in the micro-doplets becomes super-saturated until the compounds, which instead are not water-soluble, are reprecipitated in the form of NCs. The size and shape of the resulting NCs can ideally be controlled by varying the thermodynamic conditions \cite{Chung2006}.
%Dynamic light scattering %(DLS, 90Plus Particle Size Analyzer, Brookhaven Instruments) 
%measurements (Figure 1c: for the Figure ask Marianna/Daniele) show that the crystals grown under our experimental conditions are more likely to have a hydrodynamic diameter of about \textbf{\SI{200(00)}{\nano\meter}, where the uncertainty is the maximum deviation in the measurements on x nominally identical sample solutions}. The measurements are nicely fitted by a log-normal distribution which is a signature of small polydispersion of the colloids in solution. \textbf{The stability of the solution has been investigated by repeating DLS correlation measurements every day for one week, observing....}

\begin{figure}[h!]
\centering
\includegraphics[width=1\textwidth]{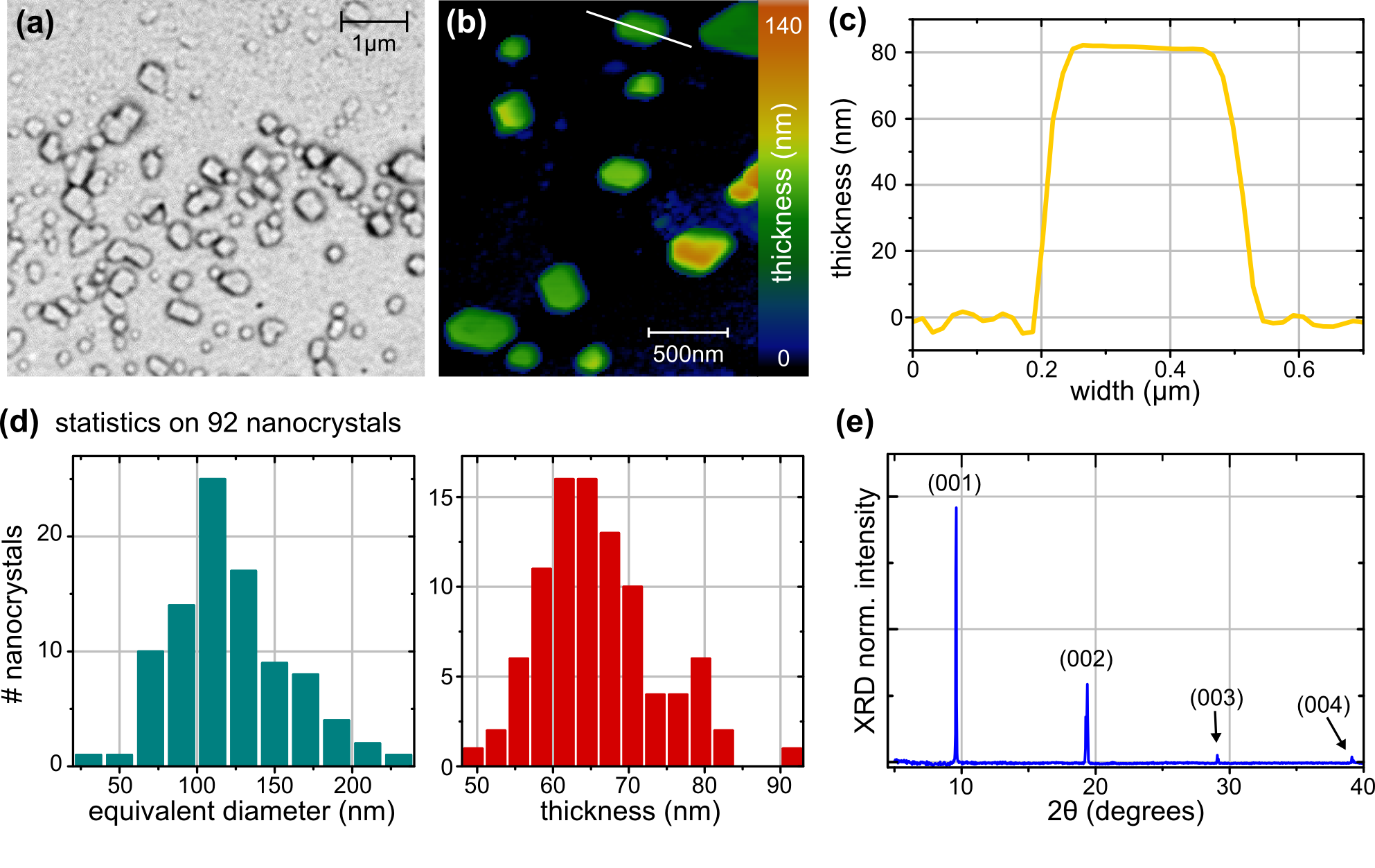}
\caption{\textbf{Morphology of DBT:Ac NCs grown via reprecipitation}: Typical SEM \textbf{(a)} and AFM topography \textbf{(b)} images. \textbf{(c)} Cross section showing the NC thickness profile along the white line in panel \textbf{(b)}. \textbf{(d)} Statistical analysis on AFM images yielding a NC equivalent diameter and standard deviation of ($113 \pm 64$)\SI{}{\nano\meter} and an average thickness of ($65 \pm 13$)\SI{}{\nano\meter}. \textbf{(e)} Normalized XRD pattern in which only the peaks from the (001) plane and higher order reflections are well resolved, due to the NCs' platelet-like morphology (with c-axis perpendicular to the substrate).}
\label{fig1}
\end{figure}

For morphological and optical characterization, a drop of the suspension of NCs in water is deposited on a coverglass substrate and dried in desiccator. Typical scanning electron microscopy (SEM) %(SEM, Phenom Pro, PhenomWorld) 
and atomic force microscopy (AFM) images are displayed in Figures \ref{fig1}a and \ref{fig1}b, respectively. For some NCs it is possible to identify peculiar features of crystalline Ac - such as the hexagonal-like morphology - while others exhibit a round-like shape, possibly due to a few \SI{}{\nano\meter} acetone-rich solvent cage. By analysing the AFM images of %performed with the open-source software Gwyddion \cite{Necas2012},
%we collect statistics
92 NCs (Figure \ref{fig1}d), we deduce an average equivalent diameter %$d_{\textup{eq}}$ 
of ($113 \pm 64$) \SI{}{\nano\meter} and an average thickness %$t$ 
of ($65 \pm 13$) \SI{}{\nano\meter}, compatible with a platelet-like shape. Such values and shape are particularly promising for the coupling to evanescent fields in proximity to surfaces \cite{Skoff2016}.
The crystalline nature suggested by the clear-cut edges and flat surfaces (see Figure \ref{fig1}c) is verified by X-ray diffraction (XRD) measurements. The XRD pattern shown in Figure \ref{fig1}e exhibits a strong diffracted peak at \SI{9.17}{\degree} - that corresponds to the (001) plane - and other equivalent periodic peaks corresponding to the (002), (003), and (004) planes, matching the crystallographic data for an anthracene monoclinic system \cite{Brock1990}. This also reveals that the Ac NCs, once deposited on the substrate, are mainly iso-oriented with the c-axis perpendicular to the substrate.
%as only the (001) and its higher order reflection are well resolved. 
Let us note that the transition dipole moment of a DBT molecule in the main insertion site of an Ac crystals is mostly oriented along the b-axis \cite{Nicolet2007b}, and thus results parallel to the substrate. 
\begin{figure}[h!]
\centering
  \includegraphics[width=1\textwidth]{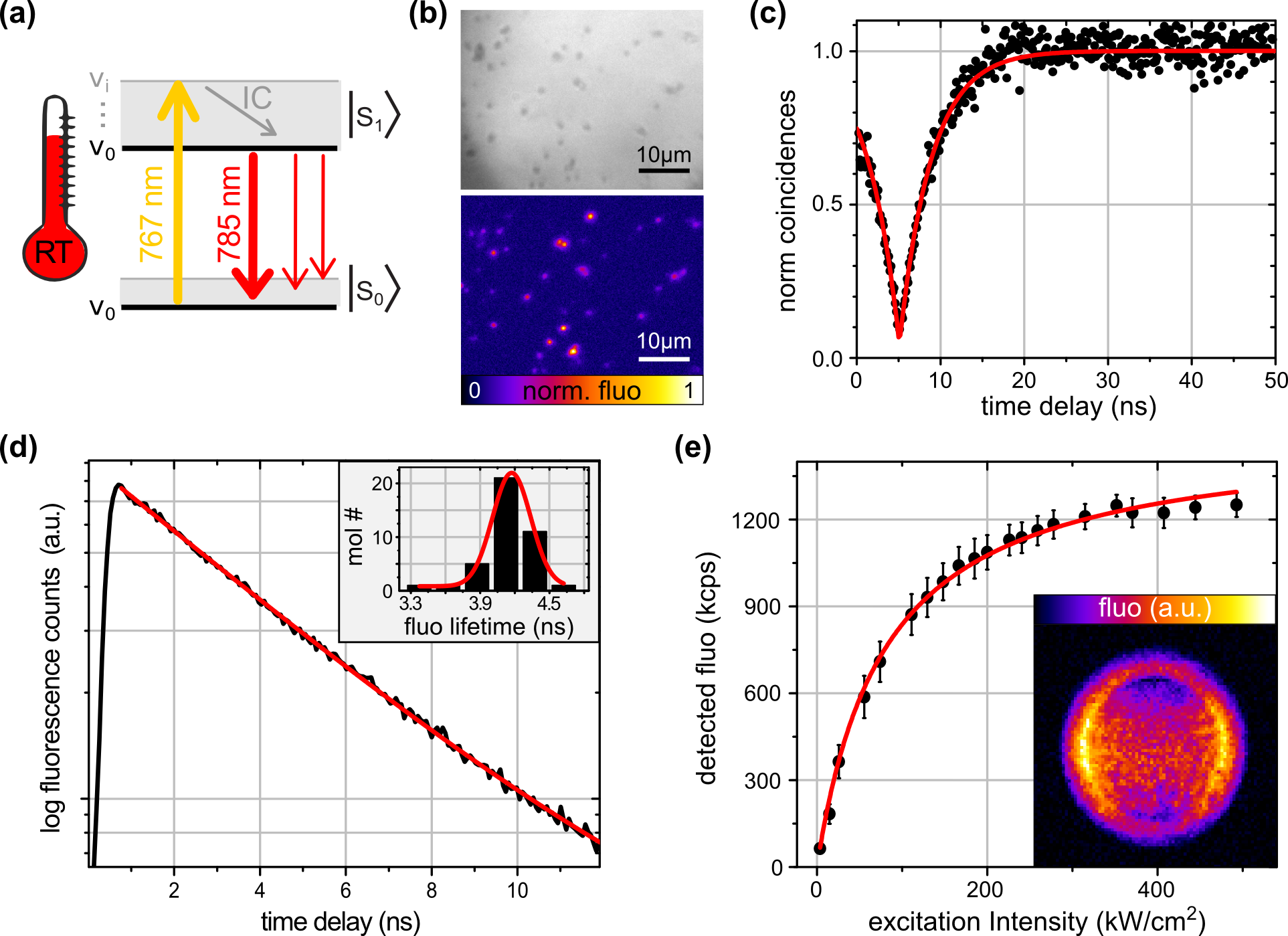}
\caption{\textbf{Photophysics of the NCs at RT}:\textbf{(a)} Off-resonant pumping scheme employed for single molecule microscopy at RT. \textbf{(b)} Comparison between white light and fluorescence wide-field EMCCD images of the same region demonstrating that about 90\% of the Ac NCs are successfully doped with DBT. \textbf{(c)} Measured photon anti-bunching ($g^2(0)=0.05$ from the fit, red solid line) from the emission of a single NC without any background correction (black dots). \textbf{(d)} Time-resolved measurement of the fluorescence decay from a single DBT:Ac NC (black line). A single exponential fit (red line) yields an excited state lifetime of $(4.2 \pm 0.1)$\SI{}{\nano\second}, as a free fitting parameter with the relative standard error. The inset shows the distribution of the excited-state lifetime collected from 40 NCs. The red curve is a Gaussian fit centered at \SI{4.2}{\nano\second} with FWHM = \SI{0.4}{\nano\second}. \textbf{(e)} Saturation measurement performed on a single DBT:Ac NC (black dots). The fitted red curve yields a saturation intensity of $I_\textup{s} = (80\pm 6)$ \SI{}{\kilo\watt\per\square\centi\metre} and a maximum detected count rate $R_{\infty}= (1.50 \pm 0.02)$ Mcps. In the inset, a typical BFP image of a single NC.}
\label{fig2}
\end{figure}

The sub-micrometric size of the crystalline matrix may compromise the optical properties of DBT molecules embedded therein, due to strain within the crystal and imperfections at the interfaces. Indeed, besides the case of quantum dots, it was reported for other nanocrystalline systems that such effects determine fluorescence instability and linewidth broadening\cite{Meltzer2001, Jantzen2016}. We thus perform single molecules microscopy and spectroscopy on DBT:Ac NCs with a home-built epifluorescence scanning confocal microscope - described in details in the Experimental Section \ref{exp} - that allows for both room and cryogenic temperature investigation. 

At room temperature (RT), the sample is illuminated with a \SI{767}{\nano\meter} continuous wave (CW) diode laser to pump DBT molecules into the vibrational band of the first singlet electronic excited state %$\ket{S_{1,\nu_i \neq 0}}$ 
(see the simplified Jablonski diagram in Figure \ref{fig2}a). After a fast (ps-timescale) non radiative relaxation process to the lowest level of the vibrational manifold, molecules decay to the electronic ground state (singlet). The resulting red-shifted fluorescence light around \SI{785}{\nano\meter} is  detected with an electron multiplied charge couple device (EMCCD). Typical white light and wide-field fluorescence images are compared in Figure \ref{fig2}b, showing that more than 90\% of the Ac NCs are successfully doped with DBT. To prove that the detected fluorescence stems from individual DBT molecules, single isolated crystals are illuminated in confocal mode with an excitation intensity of \SI{15}{\kilo\watt\per\square\centi\metre} (well below saturation) and the correlation between photon arrival times is measured with the Hanbury Brown-Twiss (HBT) setup (see the Experimental Section). Figure \ref{fig2}c shows the histogram of the observed coincidences from a single NC, featuring a strong antibunching dip. The experimental data is fitted at short time delays $\tau$ with the function $g^{2}(\tau)=1-b\cdot \exp(-|\tau|/\Delta t)$, where $\Delta t$ accounts for the excitation and spontaneous emission rates \cite{Trebbia2009} and $b$, the dip depth, is found to be $(95\pm 1)$\%. Among 40 analyzed NCs, 73\% of them displays an antibunching dip larger than 50\%, demonstrating that the proposed recipe is reliable to grow individual Ac NCs in 2/3 of cases doped with single DBT molecules. The purity of this system, \textit{i.e.} the second-order correlation function at zero time delay $g^{2}(0)$, can be as low as $0.05 \pm 0.01$ without any background correction.% \textbf{ADD HERE DISCUSSION ON CORRECTION.} From the statics of the photon arrival times, we can reconstruct the $g_{2}$ function also at long $\tau$, hence 

To gain further information on the emitter properties, we study the relaxation dynamics by means of time-correlated single-photon counting (TCSPC) measurements, collecting photons emitted after Ti:Sa-pulsed excitation (average intensity equal to \SI{20}{\kilo\watt\per\square\centi\metre}) with a single-photon avalanche diode (SPAD). Figure \ref{fig2}d shows a typical measured fluorescence decay curve from which the lifetime $\tau_{f}$ of the excited state can be derived via a single exponential fit in the presence of a constant background. The fit (red curve) yields an excited state lifetime of $(4.2 \pm 0.1)$\SI{}{\nano\second}. Repeating the measurement on 40 NCs we obtain the distribution for the excited-state lifetime shown in the inset of Figure \ref{fig2}d, which can be fitted with a Gaussian centered at \SI{4.2}{\nano\second} with full width at half maximum (FWHM) of \SI{0.4}{\nano\second}, in agreement with previous studies on the bulk system \cite{Toninelli2010a, Mazzamuto2014, Polisseni2016}.
%\textbf{here add blinking..}

The brightness of the NC-based single-photon source is quantified by studying the saturation behavior of the system, non-resonantly pumped with the \SI{767}{\nano\meter}-CW laser. Measurements are performed at different excitation intensities, scanning the sample under the confocal laser spot in the small region where the NC is located and detecting the red-shifted fluorescence with a single SPAD. From the obtained fluorescence maps the mean value within an area around the brightest pixel is extracted and corrected for the background counts, which is the mean value within an area out of the NC and is linear with the laser power. Data are plotted as a function of the laser intensity $I$ (black dots in Figure \ref{fig2}e) and fitted with the function describing the saturation of the photon detection rate $R(I)$ \cite{Moerner2003}:
\begin{equation}\label{sat}
R(I) = R_{\infty}\frac{I}{I+I_s}
\end{equation}
with $I_s$ the saturation intensity and $R_{\infty}$ the maximum detected count rate. For the molecule reported in Figure \ref{fig2}e the fit-procedure yields as free fitting parameters with the relative standard errors $I_s = (80\pm 6)$ \SI{}{\kilo\watt\per\square\centi\metre} 
%values that show a great variability between different molecules, as the excitation efficiency greatly depends on the molecule orientation within the NC. 
and a maximum detected count rate $R_{\infty}= (1.50 \pm 0.02)$ Mcps. These can be considered as typical values. Accounting for the quantum efficiency of the SPAD, $\eta_{det}=50\%$, the measured $R_{\infty}$ corresponds to a collected photon rate of \SI{3}{\mega\hertz} at the detector.
%, consistent with what reported for the bulky system \textbf{ref? or we mention our result?}
Moreover, comparing this value with the theoretical one of \SI{240}{\mega\hertz} related to the measured lifetime through the relation $R_{\infty}\simeq (1/\tau_{f})$ and assuming unitary quantum yield, we estimate a total collection efficiency of our setup at RT to be around 1\%, ascribed to the limited numerical aperture of the optics and their transmission combined with the molecule emission profile \cite{Checcucci2016}.

In order to determine the alignment of DBT molecules within the Ac NCs,
%single NCs are illuminated in confocal configuration with the \SI{767}{\nano\meter}-CW laser at \SI{20}{\kilo\watt\per\square\centi\metre} and
the emission from single molecules is detected by imaging the objective back focal plane (BFP) from which the angular radiation pattern can be deduced \cite{Lieb2004}. A typical BFP image is shown in the inset of Figure \ref{fig2}e, where the emission pattern features two side lobes facing each other beyond the critical angle, corresponding to the coupling between the evanescent wave in air with the propagative wave in the coverglass. The geometry and the direction of the two lobes confirms a horizontally aligned molecule, compatible with the XRD observations.

To conclude on the observed photophysical properties of the DBT:Ac NCs at RT, let us note that the repeated excitation of the same molecule to study its saturation behavior is a qualitative proof of the stability of its fluorescence. After several hours of measurements at RT, though, some molecules start exhibiting fluorescence blinking behavior, typically before they stop to fluoresce completely. This so-called photobleaching is most probably due to chemical reactions of the dye molecule with ambient oxygen \cite{Kozankiewicz2014}, a process that is more likely to occur in conjunction with the sublimation of Ac at RT. For sub-micrometric crystals we observe that sublimation at RT takes place on a time-scale of about one day but it is completely suppressed when covering the sample with a thin layer of a water-soluble polymer, such as poly(vinyl alcohol). 
%The protective film can thus be used to enhance the photostability of DBT molecules in the nanocrystalline matrix. 
% change the optical properties of the system \cite{Khalid2015} and may also 
\begin{figure*}[h!]
\centering
  \includegraphics[width=0.8\textwidth]{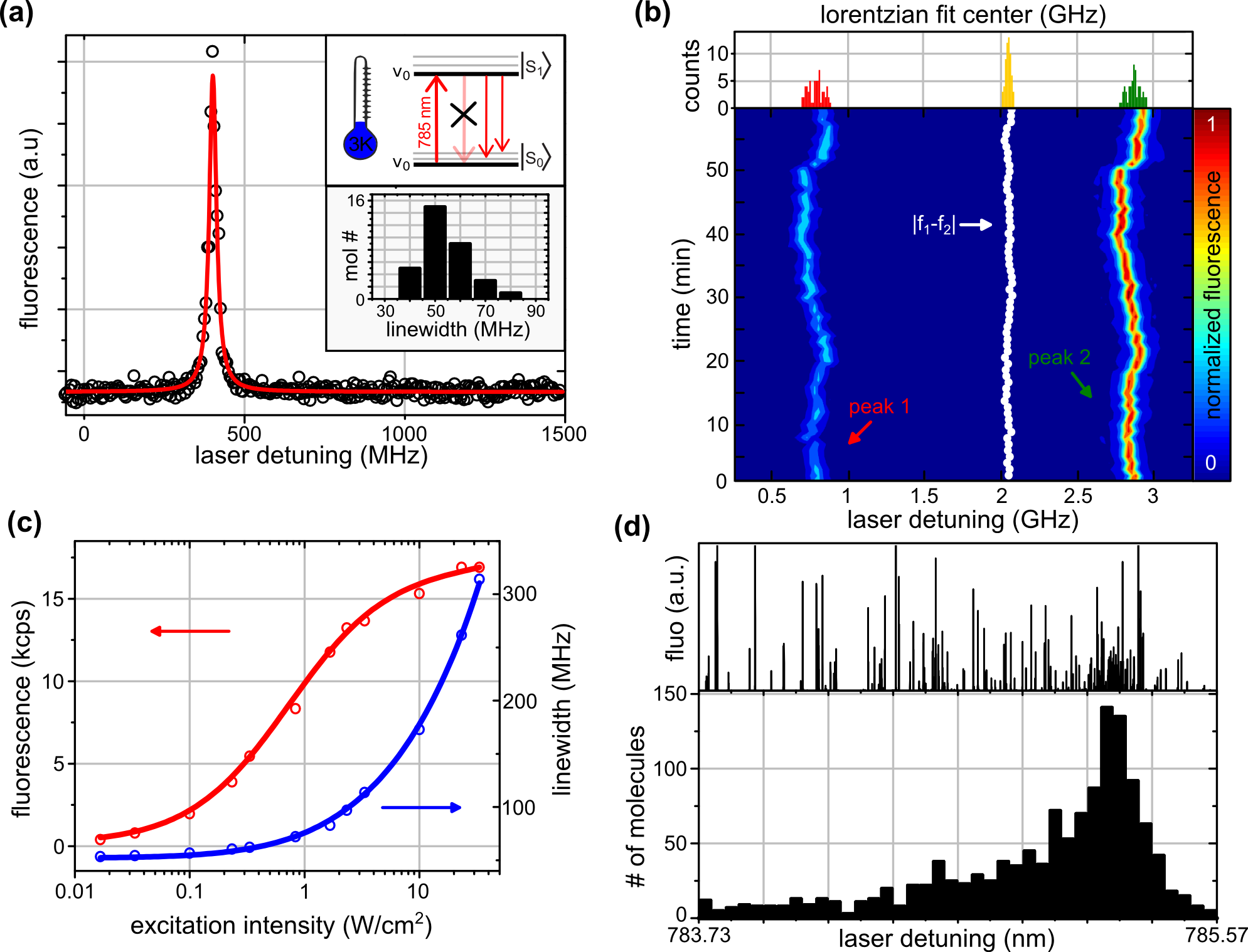}
\caption{\textbf{Photophysics of the NCs at \SI{2.9}{K}}: \textbf{(a)} Resonant excitation spectrum of a single DBT molecule in an Ac NC. The pumping scheme is sketched in the top-inset. Data (black circles) are fitted with a Lorentzian profile (red curve) and an average over four consecutive measurements yields a FWHM $=(51 \pm 10)$ \SI{}{\mega\hertz}. The bottom-inset shows the linewidth distribution of 35 molecules. \textbf{(b)} 2D plot of the excitation spectrum in time, in a frequency range where two molecules within the same NC are excited. The difference of the two peak central frequencies is plotted as white circles in the map, while the relative distributions are shown as histograms in the top panel. \textbf{(c)} Saturation curve (red circles) and power broadening (blue circles) of the ZPL are displayed with the theoretical fits (solid lines), yielding a maximum number of detected photons $R_{\infty}=(16.8 \pm 0.4)$ kcps. \textbf{(d)-top} Excitation spectrum collected from a single NC within a frequency range of \SI{800}{\giga\hertz} around \SI{784.6}{\nano\meter}. \textbf{(d)-bottom} Inhomogeneous distribution of the ZPLs collected from 20 NCs.}
\label{fig3}
\end{figure*}

At cryogenic temperatures, highly doped DBT:Ac NCs are studied under resonant excitation of the so-called 00-Zero Phonon Line (ZPL) ($\ket{S_{0,\nu_0}} \rightarrow \ket{S_{1, \nu_0}}$). In this pumping scheme, sketched as an inset of Figure \ref{fig3}a, single DBT molecules can be addressed spectrally one at a time by tuning the frequency of a narrow-band laser and exploiting the inhomogeneous distribution of the molecular resonances. In fact, depending on the host matrix, the ZPLs of PAH molecules can be distributed over a frequency range that can be smaller than \SI{1}{\giga\hertz} in unstressed sublimated crystals and as high as \SI{10}{\tera\hertz} in polymers or amorphous materials \cite{Veerman1999, Kramer2002, Kozankiewicz1994}. We found that a DBT concentration about six orders of magnitude higher than the one proposed for RT characterization allows to spectrally select single molecules within our experimental full range of about \SI{800}{\giga\hertz} around \SI{784.6}{\nano\meter}. This will be discussed further on in the manuscript.

Figure \ref{fig3}a shows the excitation spectrum of a single DBT molecule illuminated in confocal mode at \SI{0.3}{\watt\per\square\centi\metre} (below saturation), recording the red-shifted fluorescence as a function of the laser frequency (black circles). The spectral line is fitted with a Lorentzian profile (red curve) yielding a FWHM of \SI{51}{\mega\hertz}, with an uncertainty of \SI{10}{\mega\hertz} given by the standard deviation of four consecutive measurements of the same spectrum. Repeating this procedure on 35 molecules in different NCs leads to the distribution displayed in the inset of Figure \ref{fig3}a, with a low-width cutoff consistent with the lifetime-limited value of \SI{40}{\mega\hertz}. The presence of molecules with broader linewidth can be explained in terms of the reduced size of the NC, which provides a less homogeneous environment for DBT molecules than that of bulk Ac. Moreover, interface effects on fluorescence stability and linewidth broadening are more likely to occur. However, let us note that the observed linewidth distribution is narrower than that measured by Gmeiner \textit{et al.} \cite{Gmeiner2016}, confirming the high crystallinity of the Ac NC grown via reprecipitation.

Spectral diffusion has so far hindered the deployment of traditional (inorganic) NCs for narrow-band applications, such as single-photon sources for quantum technologies. We hence carefully analyze the spectral stability of the molecule transition frequency. In Figure \ref{fig3}b, the NC fluorescence counts detected from a SPAD are displayed in a 2D color map, obtained repeatedly scanning for 1 hour the excitation frequency of the pump laser over \SI{10}{\giga\hertz}. 
The excitation of two different molecules can be recognized. The mean values and standard deviations for the two molecule ZPL frequencies over all measurements are $(65 \pm 6)$\SI{}{\mega\hertz} for peak 1 and $(59 \pm 4)$\SI{}{\mega\hertz} for peak 2. However, the common-mode fluctuation of the peak central frequencies is a clear indication of the non-negligible contribution given by the pump laser instability (the laser diode is thermally stabilized but not referred to any absolute frequency standard). To get rid of this contribution and highlight possible spectral diffusion we analyze the distribution of the the two peak central frequencies difference (see the top panel in Figure \ref{fig3}b), plotted as white circles in the map. The maximum variation of such differential value is \SI{17}{\mega\hertz}, which is well within the molecule linewidth and suggests negligible spectral diffusion for DBT:Ac NCs at \SI{3}{\kelvin}. The same analysis has been carried out on couples of molecules in 8 different NCs. We observed sizable fluctuations only in one case where the ZPL central frequency over time exhibited a standard deviation of \SI{54}{\mega\hertz}.
%In this case, from the time traces of the excitation spectra of the 2 molecules %we calculate FWHM = $64 \pm 7$ \SI{}{\mega\hertz} for peak 1 and FWHM = $86 \pm 30$\SI{}{\mega\hertz} for peak 2. 
%we calculate a maximum variation of the difference of the two peak centers of \SI{54}{\mega\hertz}, which thus represents an estimation of the spectral diffusion. 
%\textbf{This value is anyhow very small with respect to......}
As a term for comparison, we remind here that aromatic molecules in polymers or other amorphous hosts present linewidths as large as few GHz, accompanied by large spectral jumps (of the order of tens of \SI{}{\giga\hertz}) \cite{Walser2009, Kozankiewicz1994, Boiron1999}. Also, when molecules are embedded in a poor crystalline environment, a broadening of both linewidth  and inhomogeneous distributions is observed, and spectral jumps, even if in a narrow frequency range of tens of \SI{}{\mega\hertz}, are more likely to occur \cite{Gmeiner2016}.

Figure \ref{fig3}c shows in logarithmic scale a typical saturation profile of a single molecule and its line broadening at low temperatures, obtained by measuring the excitation spectrum for several pump powers and plotting the detected count rates at resonance (blue circles) and the FWHMs (red circles) as a function of the excitation intensity. Detected counts are fitted with equation \ref{sat}, providing a saturation intensity $I_s=(0.73 \pm 0.03)$\SI{}{\watt\per\square\centi\metre} and a maximum number of detected photons $R_{\infty}=(16.8 \pm 0.4)$kcps (free fitting parameters with the relative standard errors), or equivalently $R_{\infty}=$ \SI{33.6}kcps accounting for the detection efficiency $\eta_{det}$. This count rate is compatible with the collection efficiency of our experimental setup for low temperature measurements of about \SI{0.3e-3}, mainly due to the the orientation of the emissive dipole and the low numerical aperture of the collecting optics. The power broadening of the homogeneous spectral line $\gamma_{hom}(I)$ fits perfectly with the expected saturation law (blue line in Figure \ref{fig3}c) given by the equation \cite{Ambrose1991}:
\begin{equation}
\gamma_{hom}(I) = \gamma_{hom}(0)\left(1+\frac{I}{I_{s}}\right)^{1/2}
\label{broadening}
\end{equation}
which assumes negligible spectral diffusion, as previously demonstrated. 

The inhomogeneous broadening of DBT molecules in Ac NCs is studied tuning the excitation laser over the available frequency range of about \SI{800}{\giga\hertz}. In the top of Figure \ref{fig3}d a typical excitation spectrum collected at \SI{2.9}{\kelvin} from a single NC is displayed, where we can distinguish about 80 peaks, each corresponding to a single molecule.
%, yielding a typical density of \textbf{1 molecule per tot MHz}. 
The same measurement performed simultaneously on 20 NCs illuminated in wide-field for two orthogonal polarization of the laser pump allows to estimate the inhomogeneous distribution of the ZPLs of DBT molecules in Ac NCs. The result of this analysis is plotted in the histogram on the bottom of Figure \ref{fig3}d. We deduce a mean value of the transition frequency equal to \SI{785.1}{\nano\meter} with a standard deviation of \SI{0.4}{\nano\meter}, which is in agreement with the inhomogeneous broadening measured for other dyes in crystalline systems \cite{Moerner2003}. Finally, we observe that DBT:Ac NCs are ideal for the deterministic integratation into nanophotonic devices, opening new perspectives on the use of molecules in the development of real-world quantum technologies.

\section{Conclusions}
In this work we demonstrate organic nanocrystals doped with quantum emitters, performing as efficient, photostable and scalable single-photon sources, at both room and cryogenic temperatures. In particular DBT:Ac crystals are presented, with an average size of few-hundreds nanometer. The growth procedure is based on reprecipitation, an inexpensive method that is adapted for a precise tuning of DBT concentration.
%show that the crystals grown under our experimental conditions are more likely to have a hydrodynamic diameter of about
Atomic force microscopy shows that the crystals grown under our experimental conditions present an average thickness of about \SI{60}{nm} and an average size of \SI{100}{nm}. The reported values can be controlled and reduced by varying the reprecipitation conditions, such as water temperature, droplet size, injected solution concentration and addition of surfactants. X-ray diffraction confirms the crystallinity of the nanoparticles and their platelet-like morphology. At room temperature, single DBT molecules in NCs show a maximum detected count rate of \SI{1.5}{\mega\hertz}, a multi-photon probability lower than 5\% and a well defined dipole orientation. At \SI{2.9}{\kelvin}, the vast majority of molecules exhibits linewidths close to the lifetime-limited value and a relative narrow inhomogeneous distribution of \SI{180}{\giga\hertz} around \SI{785}{\nano\metre}. Accurate investigation on their photostability demonstrates that each NC embeds several molecules with stable fluorescence lines, with no signs of blinking or spectral diffusion on time scales of hours. These results may be extended to different molecular host-guest systems, functionalization protocols and purposes, making active organic nanocrystals a new toolbox for the integration of quantum emitters in photonic and optoelectronic circuits, as well as in complex hybrid devices. 
%Overall, we expect this work to open the pathway for (organic?) nanocrystals to quantum optics applications as well as ....
%Indeed, the possibility to match molecule transition frequencies by applying a Stark voltage (Kolchevko2009), pave the way towards the deterministic implementation of several single photon sources in parallel to realize complex photonic networks for on-chip quantum operations.
%Overall, we believe the proposed novel fabrication method results in an improvement in the performance of molecule-based system as single photon source and further engineering efforts in their integration in writable polymeric structures may facilitate the transition of molecules from a proof-of-concept to practical realistic application. 

\section{Experimental Section}\label{exp}
\textbf{DBT:Ac NCs growth protocol.} The DBT:Ac NCs growth procedure consists in injecting \SI{250}{\micro\liter} of a mixture $1:10^6$ of 1mM DBT-toluene and 5mM Ac-acetone solutions into \SI{5}{\milli\liter} water. While continuously sonicating the system for \SI{30}{\minute}, solvents dissolved in water and DBT:Ac crystals are formed in aqueous suspension. Solvents and Ac are purchased from Sigma Aldrich, water is deionized by a Milli-Q Advantage A10 System (\SI{18.2}{\milli\ohm\cm} at \SI{25}{\celsius}) and DBT is purchased from Mercachem.\\
\textbf{Morphological characterization.} Crystals size is evaluated by scanning electron microscopy (SEM, Phenom Pro, PhenomWorld) and atomic force microscopy (AFM, Pico SPM from Molecular Imaging in AC mode equipped with a silicon probe NSG01 (NT-MDT) with \SI{210}{\kilo\hertz} resonant frequency). XRD measurements were performed at CRIST, the Crystallographic Centre of the University of Florence (Italy), with a XRD Bruker New D8 on a sample made of few \SI{}{\micro\liter} of suspension desiccated on a silicon-low background sample holder (Bruker AXS).\\
\textbf{Optical setup.} The optical characterization of DBT molecules within the sub-um Ac crystalline matrix was performed with a versatile home-built scanning fluorescence confocal microscope. The setup is equipped with a closed cycle Helium cryostat (Cryostation by Montana Instruments), capable of cooling samples down to \SI{2.9}{\kelvin}. Molecules can be excited at \SI{767}{\nano\meter} either by a continuous wave laser (CW, Toptica DL110-DFB) and a pulsed Ti:Sapphire (\SI{200}{\femto\second} pulse width, \SI{81.2}{\mega\hertz} repetition rate) laser. Alternatively, at cryogenic temperature, resonant excitation is performed with a narrowband fiber-coupled CW laser (Toptica, LD-0785-0080-DFB-1) centered at \SI{784.6}{\nano\meter}, whose frequency can be scanned continuously over a range of \SI{800}{\giga\hertz}. All laser sources are linearly polarized to allow optimal coupling to single DBT transition by means of a half-wave plate in the excitation path. The laser intensities reported in the main text are calculated from the power measured at the objective entrance divided by the area of the confocal spot measured on the bare substrate (in both cases larger than the diffraction limited spot).
For low temperature measurements, the excitation light is focused onto the sample by a long working distance air objective (Mitutoyo $100\times$ Plan Apochromat, NA $= 0.7$, WD = \SI{6}{\milli\meter}) and can be scanned over the sample %of \SI{}{\micro\metre\square} 
through a telecentric system and a dual axis galvo-mirror.
For room temperature measurements, a high-NA oil immersion objective (Zeiss Plan Apochromat, $100\times$, NA$=1.4$) is used to focus light on the sample which is mounted on a piezoelectric nanopositioner (NanoCube by Physik Instrumente). 
The Stokes-shifted fluorescence is collected by the same microscope objective used in excitation, separated from the excitation light through a dichroic mirror (Semrock FF776-Di01) and a longpass filter (Semrock RazorEdge 785RS-25) and detected by either an EM-CCD camera (Andor iXon 885, $1004\times1002$ pixels, pixel size \SI{8}{\micro\meter}$\times$\SI{8}{\micro\meter}) 
%equipped with a spectrometer (Andor Shamrock, 0.1nm resolution) 
or two single-photon avalanche diodes ($\tau$-SPAD-50 Single Photon Counting Modules by PicoQuant). SPADs can be used independently or in a Hanbury Brown-Twiss (HBT) configuration, using a time-correlated single-photon counting (TCSPC) card (PicoHarp, PicoQuant). A converging lens can be inserted in the excitation path to switch between confocal and wide-field illumination while a converging lens is added in the detection path before the EMCCD camera to study the wave-vector distribution of the light emitted by single DBT molecules via BFP imaging. 

\section{Notes}
The authors declare no competing financial interest.

\begin{acknowledgement}
The authors would like to thank S. Ciattini, L. Chelazzi (CRIST) for helping with XRD measurements, M. Mamusa for dynamic light scattering experiments, F. Intonti for the microinfiltration setup, D. S. Wiersma for access to clean room facilities, M. Bellini and C. Corsi for Ti:sapphire operation, K.G. Sch\"{a}dler and F.H.L. Koppens for helpful feedback on the NCs properties and useful discussions about integration in hybrid devices. This work benefited from the COST Action MP1403 (Nanoscale Quantum Optics). The authors acknowledge financial support from the Fondazione Cassa di Risparmio di Firenze (GRANCASSA) and MIUR program Q-Sec Ground Space Communications.

\end{acknowledgement}

\bibliography{NCpaper}

\end{document}